**Visible-Light Photoionization of Aromatic Molecules in Water-Ice: Organic Chemistry across the Universe with Less Energy**

Antti Lignell[1,2,*], Laura I. Tanalenda-Ossorio[1,4], and Murthy S. Gudipati[1,3,*]

* Corresponding authors: AL Tel: +358-505911666; E-mail: lignell@gmail.com
MSG Tel: +1-818-3542637; E-mail: murthy.gudipati@jpl.nasa.gov

[1] Ice Spectroscopy Lab, Science Division, Jet Propulsion Laboratory, California Institute of Technology, 4800 Oak Grove Drive, Pasadena, California 91109, USA.

[2] Present address: Department of Chemistry, P.O. Box 55, 00014, University of Helsinki, Finland

[3] Institute for Physical Science and Technology, University of Maryland, College Park, MD 20742, USA.

[4] Present address: Universidad EAFIT, Carrera 49 # 7 sur 50, Medellín, Antioquia, Colombia.

**Abstract**

Ionization of gas-phase organic molecules such as polycyclic aromatic hydrocarbons (PAHs) requires vacuum ultraviolet photons at wavelengths shorter than 200 nm (~6-9 eV). We present here for the first time that visible photons - accessible through sunlight - can cause photoionization of trapped PAHs in cryogenic water-ice, accounting for 4.4 eV less ionization energy than in the gas phase. This finding opens up new reaction pathways involving low-energy ionization in many environments where water and organic matter coexist. This include the interstellar medium, molecular clouds, protoplanetary disks, and planetary surfaces and atmospheres (including Earth).

**Graphical Abstract**

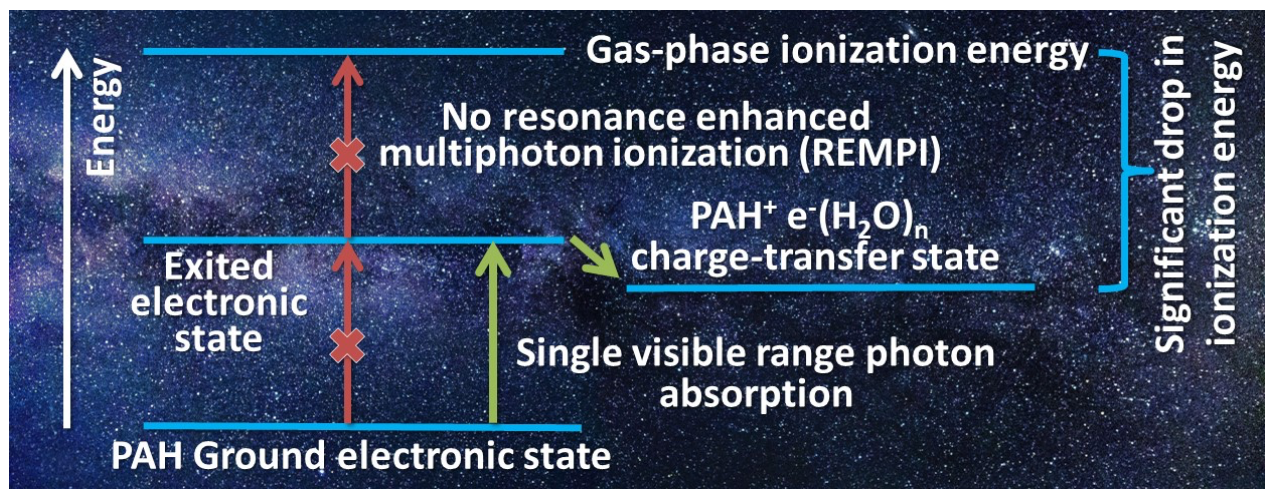

**Introduction**

Evolution of organic molecules in icy regions in the universe[1, 2] is key to understanding the origin and evolution of life on Earth[3] and the formation and dynamics of our solar system, including icy bodies such as moons of giant planets,[4] comets,[5] and meteorites.[6] Interstellar matter, the reservoir of material forming new solar systems where water and simple organics coexist,[7] is also a place where complex chemical reactions take place. These mixed ices containing frozen organics and water are also subjected to stellar radiation, initiating a wide variety of radiation-induced chemistry in these ices leading to the production of complex organic molecules, including the building blocks of life such as amino acids[8] and nucleotides.[9] Close to Earth, organics exist naturally and as anthropogenic pollutants, mixed with fog[10] or ice in snowpacks,[11] in the cryosphere,[12] and in atmospheric clouds[13] as well as heterocyclic aromatic backbones of DNA, RNA, and proteins. These organics in ices are exposed to solar photons leading to complex photochemical transformations. Thus, the co-existence of water and organics in the form of ice particles is a universal phenomenon and their evolution under radiation manifests itself in research topics ranging from astrophysics through the origin of life to atmospheric sciences.

Radiation (photons, electrons, cosmic rays, ions, and solar wind) bombardment of ice was assumed to result only in molecular dissociation and radical formation within these ices until recent work showed that ionization and ionization-mediated chemistry of polycyclic aromatic hydrocarbons (PAHs) in ice is prevalent.[14, 15] This included initial indications that ionization energy of PAHs in ice may be lower than in the gas phase[16] and that is also corroborated with such a trend observed in theoretical works.[17] Lowering ionization energy is also observed for DNA nucleobases in aqueous medium.[18] However, all these studies indicated about 2 eV lowering of ionization energy compared to the gas-phase values. In contrast, our study presented here demonstrates for the first time that photons at the visible wavelengths representing sunlight reaching Earth's surface could trigger photoionization of organic molecules trapped in ice, causing an unprecedented lowering of the ionization energy of organic molecules by as much as 4.4 eV. Our studies open up new pathways to understand chemical transformations of organic matter in water-ice under radiation with significant implications to a wide range of phenomena that are discussed below.

**Results**

**Photoionization with Tunable Wavelength**

We vapor deposited a PAH/water film with a typical ratio of 1:1000 that produced a monomeric sample. This matrix-ratio has long been demonstrated to yield a well-isolated species. Monomeric sample isolation is important since we wanted to exclude further complications from heterogenic samples to get quantitative results for the production of monomeric radical cation well below the gas phase ionization (IP) potential of a corresponding neutral counterpart. These gas phase IPs are 7.43, 6.97, and 6.96 eV for pyrene, tetracene, and perylene, respectively.[19]

Using a defocused laser beam from a tunable optical parametric oscillator (OPO) laser that could be operated between 210 nm and 700 nm, we conducted a systematic study of the ionization of PAH molecules trapped in amorphous water ice. We found that trapped PAHs studied here: pyrene, tetracene, and perylene, are ionized at energies 3.7 eV, 2.6 eV, and 2.88 eV, respectively. The maximum shift was found for tetracene at ~4.4 eV below the gas phase ionization energy. The corresponding wavelength of the photon is shifted from 177 nm (vacuum ultraviolet) to 475 nm that is in the visible range (blue), relatively close to the solar emission maximum at ~500 nm.[20] Further studies showed that ionization is efficient when the OPO laser wavelength is tuned to the electronic absorption bands, as shown in **Figure 1**, indicating that the primary excitation occurs into the valence states of the trapped organic species. These absorptions are collected into **Table 1**. Data shown in the **Supplementary Figure S5** further demonstrates this phenomenon.

While the majority of the work reported here is focused on amorphous water ices at low- temperatures, we have conducted preliminary studies that show similar low energy photoionization of PAHs in crystalline ices (when amorphous ices containing PAHs were warmed up to 140-150 K temperatures to induce phase transition[21] and cooled down back to 5 K). However, as detailed in the Materials and Methods section, aromatic molecules tend to form aggregates in crystalline ice,[21] which dictate the ionization efficiency.[15] Polar or small organic molecules are likely to be trapped at higher dilution in crystalline ice (including the formation of clathrates) and expected to behave similar to isolated organic chromophores in amorphous ice. Further studies need to be conducted to fully explore the range of conditions under which photoionization with lower energy photons dominates their chemistry in a wide range of environments – from radiation biology on Earth to the evolution of organic matter in the icy environments of the Universe. An important opening of such direction is a recent work by Noble *et al.* on PAH photochemistry in different ice structures.[22]

**Excluding Multi-photon Ionization**

Due to a possibility of multi-photon ionization under pulsed laser irradiation, it was important to show that our results were due to an absorption of a single photon since multiphoton ionization is significantly less likely to happen in nature. We did confirm this with three independent ways shown below.

First, as shown in **Figure 2** (and **Supplementary Figure S6)**, the photo bleaching ratios followed single exponential curve. When the laser power (photon flux) is ramped up, the increased number of photons results to the proportional (linear) dependence of ionization rate as a function of number of photons and that is very characteristic to a single-photon excitation. Two-photon process would have yielded a quadratic dependence between the ionization rate and the photon flux. The **Supplementary Figure S6** clearly demonstrates that the photolysis is following a single exponential curve when the four different laser fluxes are converted to the corresponding time units. Second, we used a 75 W Xe-arc lamp instead of the nanosecond pulsed OPO laser for the photolysis (further details in Materials and Methods). We used broadband filters to match the absorption maxima of PAHs. The photon flux from the lamp is significantly lower ($\sim 10^{12}$ photons/cm$^2$/nm) than from the tunable pulsed OPO and unquestionably cannot lead to the multi-photon ionization. The lamp irradiation also resulted to the ionization of the PAH as shown in **Supplementary Figures S7 and S8**. Finally, we did a control experiment where inert argon matrix was used as a host for PAH tetracene, using the same Xe-arc lamp as mentioned above under identical conditions, which did not yield any detectable ionization of PAHs (**Supplementary Figure S9**) indicating that the water matrix is critical to effectively ionize PAHs. All of these three experiments confirm that the ionizations in our PAH/water samples were indeed due to the single-photon process and could be occurring under natural environment in condition where the light from the nearby stars is illuminating a PAH inside water ice.

**Charge-transfer State Mediated Mechanism**

Our findings can be rationalized based on the recent advancements in understanding of hydrated electron in water clusters and by invoking charge-transfer transitions in ice. The vertical binding energies (VBE) of electrons with water clusters are known[23] to be significant ($\sim 3.2$ eV for inside the cluster and $>3.5$ eV on the surface), meaning that water-ice has a high electron affinity and the electron affinity varies with water cluster-size. Thus, PAH$^+$ e$^-$(H$_2$O)$_n$ charge-transfer states could exist in water-ice that are accessible at much lower energies than the gas-phase ionization energies of 7.43 eV (pyrene), 6.97 eV (tetracene), and 6.96 eV (perylene).[19] Our earlier work on noble-gas oxides (NgO) has shown that such empirical formulation yields quantitative estimates.[24] We conducted similar analysis for PAH-$(H_2O)_n$ charge-transfer states, where $(H_2O)_n$ represents bulk amorphous ice. Further details are given in the Materials and Methods Section. From these calculations, based on the distance between the charge centers, a maximum lowering of ~5 eV could be expected for the ionization of aromatic molecules in the ice compared to the gas phase values (please see above), that is in line with our experimental observations. It is important to note here that a Coulomb interaction between the positive ion and its negative counter-part (electron trapped in ice) increases with decreasing ion-pair distance R. Hence smaller the ionized atom/molecule is, greater would be the reduction of its effective ionization energy. As detailed in the Materials and Methods Section, H and $H_2$ are expected to show the largest shifts in ionization energy from the gas phase to trapped in water-ice.[25]

Further investigations revealed that ionization reaches a plateau at a given wavelength and starts again at a higher energy (shorter wavelength) indicating that these charge-transfer states are accessible at different energies for different PAH-trapping sites in ice, in agreement with significant heterogeneous broadening of PAH absorption bands in ice.[14] Energies of charge-transfer states are controlled by the local cluster size of water molecules forming the cavity in which a PAH molecule is trapped. Larger clusters have higher electron-binding energy. When tuned to the lowest energy (longest wavelength) of the PAH electronic absorption band (pyrene at 335 nm, tetracene at 475 nm, and perylene at 430 nm), photoionization is saturated after few minutes of irradiation and no significant change is seen under prolonged (>2h) irradiation, indicating that PAHs in these sites are fully ionized. When the laser wavelength is shifted to match with the next higher energy absorption band (e.g. 470 nm → 440 nm with tetracene), the photoionization can reach another plateau, as schematically shown in **Figure 3** (also see **Supplementary Figures S10 and S11**). Thus, our work presented here shows that ionization of aromatic molecules trapped in ice is mediated by the valence excitations that cross into charge-transfer states, and that the photoionization is mediated by one-photon absorption in ice for a wide wavelength range, starting with the lowest electronic excited states of the organic molecules at 335 nm (pyrene), 430 nm (perylene), and 475 nm (tetracene).[26-28] The electronic transitions of three model PAHs that we used in this article, have been extensively studied and the lowest energy abortions that are visible in spectra represent $S_0$-$S_1$ transitions for perylene and tetracene, and $S_0$-$S_2$ for pyrene.[27, 29, 30] The small oscillator strength of pyrene $S_0$-$S_1$ transition (<1% compared to $S_0$-$S_2$) makes this absorption spectroscopically "dark" and thus our lowest energy excitation was targeting the second electronic excitation state at 335 nm.[29]

Multiphoton Ionization, whether resonance enhanced (REMPI) or non-resonant was ruled out by three independent ways mentioned earlier. A simplified potential energy diagram of this situation is presented in **Figure 3.** At shorter photolysis wavelengths, where the electronic absorption spectra of ionized aromatic molecules and their neutral parent molecules overlap, both the parents and ions are further ionized or subjected to secondary photochemical processes. Consequently, both neutral and ion absorption bands decrease with increasing shorter- wavelength irradiation.

**Thermal Electron-Ion Recombination**

During the photoionization of PAHs inside an ice cavity, the free electrons will be trapped into the surrounding ice matrix. So far there is no direct detection of electrons in ice through transient or static spectroscopic methods. We observed indirect evidence for "trapped or loosely bound electrons" in ice during thermal warm-up of ice and through photochemical activation of the trapped radical cations. In both cases, as shown in **Figure 4**, a part of the radical cation absorption decreased with simultaneous increase in the neutral parent molecule absorption, indicating electron-ion-recombination. These experiments indirectly confirm the availability of loosely bound electrons trapped in ice. This observation has wide reaching consequences that are discussed below, showing for the first time availability of electrons in ice.

**Discussion**

Our work presented here establishes that organic molecules trapped in water-ice can be ionized at significantly lower energies, making such photoionization processes accessible over much wider environments in the Universe (please see Introduction for details).[14, 15] Once ionized, the produced radical cations are highly reactive and they could react with surrounding species in the ice, leading to the evolution of complex organics in these environments. Some of the photoproducts being biologically important molecules, for example PAHs are shown to be hydrogenated, oxygenated, forming hydroxy and ketone functionalized PAHs in ice upon radiation processing using both in-situ and subsequent room-temperature analyses.[31, 32]

The studies presented here have implications from astrophysics to atmospheric chemistry. Dense molecular clouds (DMCs) in the interstellar medium are known to be the birthplaces of new stars and solar systems. So far, it is assumed that the photochemistry in these DMCs, which are very cold (~10 K) and harbor the highest density of ice grains containing organics, can only occur on their outer edges.[33] If an average of 8 eV photon (~156 nm) is needed to ionize an atom or a molecule in the gas phase, these atoms/molecules could be ionized when trapped in ice at ~3.5 eV (~350 nm). Larger aromatic molecules such the ones studied here could be ionized with visible-range photons in ice, making such ionization processes accessible in the interiors of DMCs as long as visible wavelength photons could reach these depths. Our research presented here shows that lower energy photons that travel further into molecular clouds could efficiently ionize organics in these ices - triggering complex chemistry even before the formation of a circumstellar disc. Similar enhanced photochemical reactions deeper into the disks of young solar objects (YSOs) could continue into an evolved solar system like ours. Charge generated due to photoionization in ice particles could result in stronger Coulomb attraction forces between the ice particles, enabling nucleation processes in the interior of DMCs that eventually lead to gravitational collapse of molecular clouds to planetary nebula. A similar process could trigger cloud formation in Earth's atmosphere.

Red and orange dwarf stars are together by far the most common star types in Milky Way and have extremely long lifetime and much lower UV flux compared to their more massive counterparts. Planetary systems consisting red and orange dwarfs can now be more seriously considered harboring chemical conditions for life, as they can utilize the lower-energy (red-shifted) visible photons from the parent star for chemical evolution of organics within icy environments potentially leading to complex prebiotic chemistry. Interestingly, many exoplanets are found orbiting red dwarf stars in the habitable zone, where the surface temperature allows water to exist in a liquid form.[34] Important examples of such exoplanets are relatively recently found Proxima b that is orbiting Proxima Centaury (the nearest star to our Sun, just 4.2 light years away)[35] and three habitable zone exoplanets orbiting Trappist-1 star about 40 light years away from us.[36]

In addition to its implications for astrophysical ices, ionization-driven chemistry can play a significant role in the formation and chemistry of atmospheric aerosols due to the vast availability of visible and near UV range photons from the solar radiation in Earth's atmosphere and on the surface.[10, 37] In the cryosphere, surface ice is exposed to visible sunlight and anthropogenic pollutants make their way into these ices, resulting in a wide variety of complex organic chemistry. The chemistry and physics of atmospheric aerosol particles have been extensively studied, including the primary and secondary organic aerosols and their ageing in the particulate matter phase,[38, 39] originating from natural to anthropogenic sources involving emissions from traffic and industry to forests.[13] These organic aerosols enhance nucleation of aqueous particles resulting in fog and cloud particles, which would make perfect candidates for solar radiation-induced ionization-driven photochemistry presented in this article.[10, 40] A better understanding of heterogeneous ice nucleation (towards the formation of cirrus clouds) enhanced by anthropogenic organic pollutants is emerging recently.[41] Our results presented here would prompt more detailed studies on the diagenesis of living or non-living organic matter trapped in cryosphere on Earth that easily receives photons with sufficient energy. Finally, inspired by our earlier work, ionization-mediated organic chemistry in ice is beginning to become a mainstream tool for synthetic and physical organic chemistry.[14, 16]

**Conclusions:**

The work presented here shows that organic matter trapped in water ice is ionized with visible wavelength photons that are prevalent throughout the Universe and reach the places that cannot be accessed by high-energy UV photons. These ionized organics in ice form the starting point for complex evolution in environments from the interstellar medium to atmospheric ice particles on Earth. Charge-mediated nucleation processes can now be understood in the light of our work that impurities trapped in ice can efficiently be ionized leading to charged ice particles even in low-energy photon environments. Once ionized, even at very cold temperatures, some of these radical cations react with water forming oxygenated and hydrogenated species, triggering complex chemical pathways in ice. Many biological molecules and aromatic pollutants strongly absorb in the visible-UV range, making these the most reactive species in ice through ionization by solar light. Coexistence of water and organic molecules is critical for life, but under radiation a delicate balance between energy-harvesting and photochemical damage would be maintained based on the environments in which these organics, life, water, and energy coexist.

Due to the fact that most of the ices on Earth exist at relatively high temperatures (>200 K) at which PAH radical cations would react with water to form complex organics, these radical cations may evade observation at these elevated temperatures in nature. Photolysis of PAHs in ice leads to oxygenation and hydrogenation of PAHs, forming alcohols, ketones, and even carboxylic acids, some of them biologically important.[31, 32] Our two-step in-situ laser ablation and laser ionization time of flight mass spectrometry (2S-LAIMS) studies indicate that prolonged irradiation with high energy photons (10.2 eV, Ly$\alpha$) leads to the formation of hydrogenated and hydroxylated PAHs,[32] also detected in other studies.[42] Circumstantial evidence shows that these reactions are mediated by ionized PAHs (radical cations) due to their efficient formation within the first few minutes of irradiation with hydrogen lamp (Ly$\alpha$) photons. Ionized PAHs are highly reactive compared to neutral PAHs, making their direct detection at higher temperatures pertinent to Earth and terrestrial planetary conditions difficult. Transient pump-probe studies with lower energy photons used in the present study, would be necessary to fully understand and explore the potential of photoionization-mediated chemistry of PAHs (and other aromatic molecules of biological importance) in water-ice to better understand their chemical evolution in the Universe.

**Materials and Methods**

All the experiments were carried out in the Jet Propulsion Laboratory's (JPL) Ice Spectroscopy Laboratory. The closed-cycle helium gas cryo-cooler (APD), capable of reaching temperatures of 5 K, was mounted into the stainless-steel high-vacuum chamber (operating pressure ~$10^{-9}$ mbar). The UV spectra were measured with fiber-optic coupled (solarization resistant) Ocean Optics Spectrometers (HR4000 or USB4000, CCD array detector) by using a deuterium/halogen-lamp dual source giving the spectral range of 210-1100nm. IR spectra of water films during the deposition and amorphous-to-crystalline transitions were monitored with a Thermo-Nicolet 6700 FTIR spectrometer and liquid nitrogen cooled MCT-B detector using 1 $cm^{-1}$ resolution and 400-8000 $cm^{-1}$ range. Example of an IR spectrum of typical experiment (pyrene trapped in amorphous water) is presend in Supplementary Figure S12. The samples were vapor deposited onto a sapphire window by letting water vapor flow through a heated deposition line sublimating PAH material onto the sample window or by using a separate oven for PAH and a vapor deposition line for water. Temperatures of the PAH material and the water vapor flow rates were carefully controlled to achieve the complete isolation of PAH molecules and the typical PAH/water ratio was 1:1000. This matrix ratio is a working compromise between the PAH signal strength and isolation. Higher PAH ratios lead to clustering of PAHs[15, 21] and introduce additional complexity to the study. Our general observation is that ionization efficiencies may vary due to clustering, but the energy required would be the same. A deposition temperature of 30 K was used to produce amorphous ice film and all the irradiation and measurements were performed at the lowest sample temperature.

The samples were irradiated with optical parameter oscillator (OPOTEK Opolette UV) with tunable range of 210-750 nm, pulse duration of 5 ns, spectral line width of 4-6 $cm^{-1}$, and a repetition rate of 20 Hz. This OPO laser beam has a nominal 3 mm diameter at the exit of the laser housing, where the pulse energy was measured. The laser is highly divergent beam with divergences of 2 mRad (vertical) and 10 mRad (horizontal). The laser beam was defocused with a f=150 mm lens or kept far away from the ice to let the divergence of the laser, resulting in an expanded laser beam at the ice to ~100 times area at the sample (UV-Vis beam area going through an ice film, ~40 mm$^2$). We estimate about two orders of magnitude reduction of photon flux on the sample. Laser power employed in this work was anywhere between 0.1 mW and 55 mW. PAH ionization was efficient even at the lowest laser power employed and no two-photon excitation statistics were detected even at the max laser power used at 55 mW. For example, at 470 nm laser wavelength (2.64 eV per photon), a 10 mW laser power at 20 Hz repetition rate would produce 0.5 mJ/pulse, which results in a laser flux density of ~10$^{15}$ photons/cm$^2$/pulse at the laser housing exit and ~10$^{13}$ photons/cm$^2$/pulse at the ice target. The photon flux densities used at the target are >10$^{11}$ and <5x10$^{13}$ photons/cm$^2$/pulse. This range of photon flux density is not sufficient to initiate any multiphoton excitations in ice – whether non-resonant or resonance-enhanced, which we also demonstrated with varying the pulse energy of the laser (**Figure 2 and Supplementary Figure S6**). Though the lifetime of excited PAHs could be in hundreds of nanoseconds,[43] the pulse duration of OPO laser is typically 5 ns and two photons need to be simultaneously absorbed within this time frame, requiring high photon flux densities. Typically, even the most sensitive two-photon (including resonant) excitation requires a photon flux density >10$^{19}$ photons/cm$^2$/pulse, at least six orders of magnitude more than what we used toconduct photoionization in ice. To achieve multiphoton processes such as the Resonance Enhanced Multiphoton Ionization (REMPI), typically lasers of a few mJ/pulse are collimated and focused close to diffraction limits to achieve required high photon flux densities. In contrast, in our experiments we defocused the beam such that only single photon processes are ensured.

Photolysis of the ice samples were also carried out using Xe-arc lamp (OBB 150 W) operating at 75 W. This lamp housing has a *f/4.5* parabolic reflector improving the photon flux at the focus by 5-6 times than the conventional lamps. The lamp was placed at a distance such that the whole ice film was illuminated. We estimate photon flux at the sample to be at the maximum ~$10^{13}$ photons/cm$^2$/s/nm, compared to similar photon flux from the OPO laser, but within a few nanoseconds of the laser pulse, making Xe-arc lamp flux to be at least 6 orders of magnitude lower than the OPO laser flux, which itself is not sufficient for multiphoton excitations. Further, we found that tetracene is photoionized while exposed to the UV-lamp that is used to take the UV spectra. This UV lamp has fiber optic cable that transmits >210 nm wavelengths and the lamp source is a combination of deuterium and halogen lamps. The output of the UV lamp is two orders of magnitude lower than the Xe-arc lamp mentioned above. Detailed studies confirm that tetracene is photoionized with the UV spectrometer (deuterium- halogen lamps) through excitation into the strongest absorption at 280 nm. Excitation into the visible absorption of tetracene by the UV spectrometer lamp mentioned above did not result in photo-ionization, mainly due to the fact that the absorption cross-section of the visible bands around 450 nm are orders of magnitude lower (**Supplementary Figure 8**) than the 280 nm absorption. As a result it would need extended periods of irradiation in the visible region to achieve similar ionization efficiencies that is beyond the present study. Thus, under the photolysis conditions used in our experiments possibility of multiphoton ionization can completely be eliminated. It is established[23] that the electron affinity (or the vertical binding energy, VBE of an electron) of water cluster can reach a maximum of 3.5 eV. In addition to the electron affinity of water, further stabilization of the charge-transfer complex could occur through Coulomb interactions dictated by the charge separation distance (R). Thus, charge-transfer excitation energy (transfer of an electron from one molecule to another) in a condensed medium is given by:[24]

$$E_{CT}^{R}(gas) = \left[ E_{IP}^{Mol} - E_{EA}^{(H_2O)n} - \frac{e^2}{R} \right] \qquad (1)$$

$$E_{CT}^{R}(ice) = E_{CT}^{R}(gas) - \frac{(\varepsilon-1)}{(2\varepsilon+1)} \frac{e^2}{R} \qquad (2)$$

where, IP and EA are ionization potential and electron affinity of an organic molecule and local ice cluster in the isolated conditions, respectively. The electron affinity of large water clusters can alone result in more than 3.5 eV reduction in ionization potential when $(H_2O)_n$ cluster is large, i.e., n>100. The term $e^2/R$ represents the Coulomb attraction, where R is the distance between the positive (after ionization) ion and negative (electron localization in the $H_2O$ cluster) ion. Here $e^2/R$ is approximated to 14.41 eV/R (Å), when $e = 1$ (full charge separation). With an approximate separation of the charge centers by ~10 Å, typical for medium-sized PAH molecules containing 10-15 carbon atoms, the $e^2/R$ term results in another ~1.5 eV reduction in the charge-transfer (CT) energy. Thus, a lowering of ~4.5 eV is reasonable to expect for aromatic molecules imbedded in large $H_2O$ clusters compared to their gas-phase ionization potentials. The second term $\frac{(\varepsilon-1)}{(2\varepsilon+1)} \frac{e^2}{R}$ in the equation (2) is the polarizability-induced stabilization of the charge in the condensed ice medium. We expect another ~0.25 eV lowering due to the second term where, $\varepsilon$ is the optical dielectric constant of the medium, which is 1.6 for ice and ~80 for liquid water. For further details, see Ref [14]. An overall lowering of up to a maximum of 5 eV ionization energy of a chromophore trapped in water-ice is thus expected from these simple, but proven empirical methods. As the separation distance between the charge centers (R) in equation (1) and (2) is proportional to the size of the molecule (or atom) in a geminate charge pair, for smaller molecules (or atoms) $e^2/R$ will be large, resulting in significant lowering of ionization energy in ice compared to larger molecules. Ideally, H atoms or $H_2$ should show the maximum shift.

**Acknowledgments**

This research was enabled through partial funding from the following NASA programs: Solar System Workings, Rosetta Science (US), Planetary Atmospheres, Cassini Data Analysis Programs, Spitzer Science Center, and Astrobiology Institute Node Early Habitable Environments (NASA Ames). JPL's DRDF and R&TD funding for infrastructure of the "Ice Spectroscopy Laboratory" is also gratefully acknowledged. This research was carried out at the Jet Propulsion Laboratory, California Institute of Technology, under a contract with the National Aeronautics and Space Administration. AL is thankful for the Academy of Finland and Caltech fellowships that funded partly this research.

**Contributions of the Authors**

MSG conceived the project. While AL did majority of the experimental work, MSG also contributed to the experimental work. LIT conducted the Xe-arc lamp irradiation experiments under the guidance of MSG. Both AL and MSG analyzed the data, wrote the manuscript, and derived the conclusions.

**Competing Interests**

The authors declare that the authors do not have competing interests.

**Data Availability**

All data used to obtain the conclusions in this paper are presented in the paper and/or the Supplementary Materials. Other data can be obtained from the authors upon request.

**Table 1.** The most prominent UV-vis absorption bands of the neutral and cationic PAHs observed in our experiemtns. As a comparison, we have collected literature data of the corresponding species in argon matrix when available.

| Pyrene | | | |
|---|---|---|---|
| neutral in H$_2$O ice | neutral in Ar matrix[44] | radical cation in H$_2$O ice | radical cation in Ar matrix[44] |
| 333 nm | 330 nm | 446 nm | 444 nm |
| 317 nm | | | |
| 271 nm | 270 nm | | |
| 260 nm | | | |
| 238 nm | 238 nm | | |
| 229 nm | | | |

| Tetracene | | | |
|---|---|---|---|
| neutral in H$_2$O ice | neutral[45] | radical cation in H$_2$O ice | radical cation in Ar matrix[46] |
| 468 nm | 456 nm | 836 nm | Several bands at 850 nm |
| 439 nm | | 725 nm | 745 nm |
| 413 nm | 398 nm | 390 nm | 392 nm |
| 270 nm | 273 nm | 346 nm | 348 nm |

| Perylene | | | |
|---|---|---|---|
| neutral in H$_2$O ice | neutral in Ar matrix[47] | radical cation in H$_2$O ice | radical cation in Ar matrix[48] |
| 430 nm | 433 nm | 534 nm | 535 nm |
| 404 nm | 405 nm | | |

# Figures

**Fig. 1. Ionization of Aromatics in Ice.** Photoionization of polycyclic aromatic hydrocarbons trapped in amorphous water ice at 5 K. Difference spectra show the result of photolysis and arrows mark the wavelengths at which the ice was irradiated. Simultaneous decrease of neutral PAH (negative bands) and increase of $PAH^+$ cation (positive bands) is observed. OPO laser wavelength is tuned to the absorption maximum of PAH molecules, which are for pyrene at 335 nm, tetracene at 475 nm, and perylene at 430 nm, corresponding to the lowering of ionization energy by ~3.5 eV, 4.4 eV, and 4.0 eV, respectively.

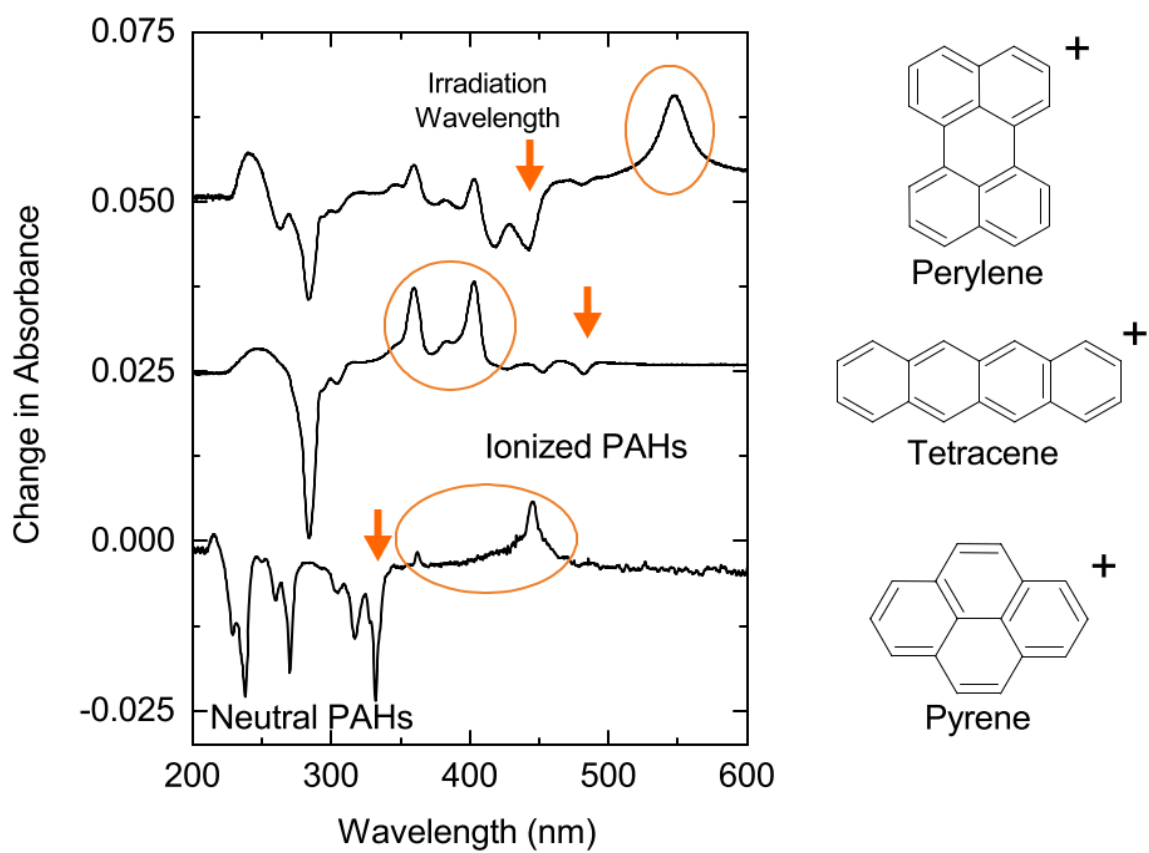

**Fig. 2. Single Photon Ionization.** Laser photoionization kinetics of tetracene in $H_2O$ ice at 5 K. Left Panel: Depletion of neutral tetracene as a function of time (dose) and increase of the radical cation absorption. Right Panel: Ruling out the multiphoton ionization. Variation of laser intensity (photon flux) by a factor of 5.5 results in an increase in photodepletion of neutral by the same factor (linear one-photon process). Open squares represent the data points from 55 mW photolysis by multiplying the time by 5.0 and plotted on the 10 mW curve. If photoionization were to require a two-photon process, kinetics at 10 mW and 55 mW should have been scaled by a factor of 25.

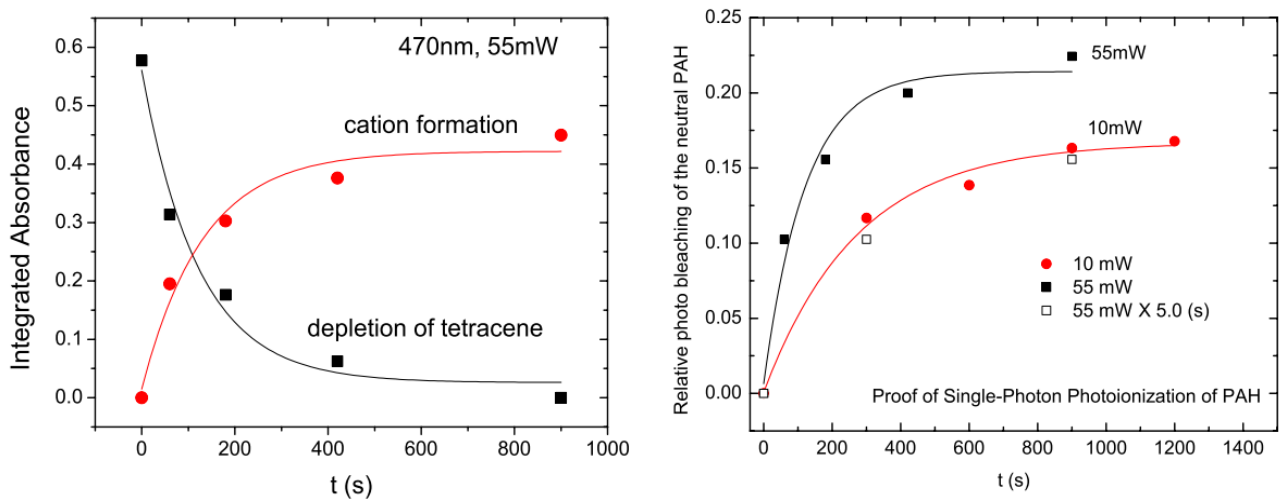

**Fig. 3. Crossing of Valence and Charge Transfer States.** Left panel: Simplified image of the potential energy surfaces during the single photon excitation/ionization. Electronic excitation of a ground state PAH ($S_0$) to the excited states ($S_1$/$S_2$) and subsequent ionization to the [PAH$^+$ (H$_2$O)$_n$e$^-$] charge transfer state (blue arrows). Resonance enhanced multiphoton ionization (REMPI, red arrows) process is ruled out with three independent ways presented in the text. Right panel: Proposed mechanism for the photoionization of strongly absorbing organic chromophores trapped in water-ice. Blue curves represent the energies of charge-transfer (CT) states at a given electron affinity of amorphous water-ice cage. Nuclear repulsions result in an increase of the CT potential at shorter distances (red curves). The CT states cross valence excited states carrying significant oscillator strength as in the case of PAHs, as shown on the left side for pyrene electronic absorption spectrum ($S_0$-$S_1$ transition being the dark singlet state absortion for pyrene).[29] Consequently, excitation into valence states results in photoionization of PAHs in ice, reducing the ionization-energy several eV lower than the gas-phase ionization energy. When photolyzed at shorter wavelengths the excess of energy both in ionized PAH (radical cation) and the electron lead to secondary reaction pathways depleting both neutral and ionized PAH populations in the ice.

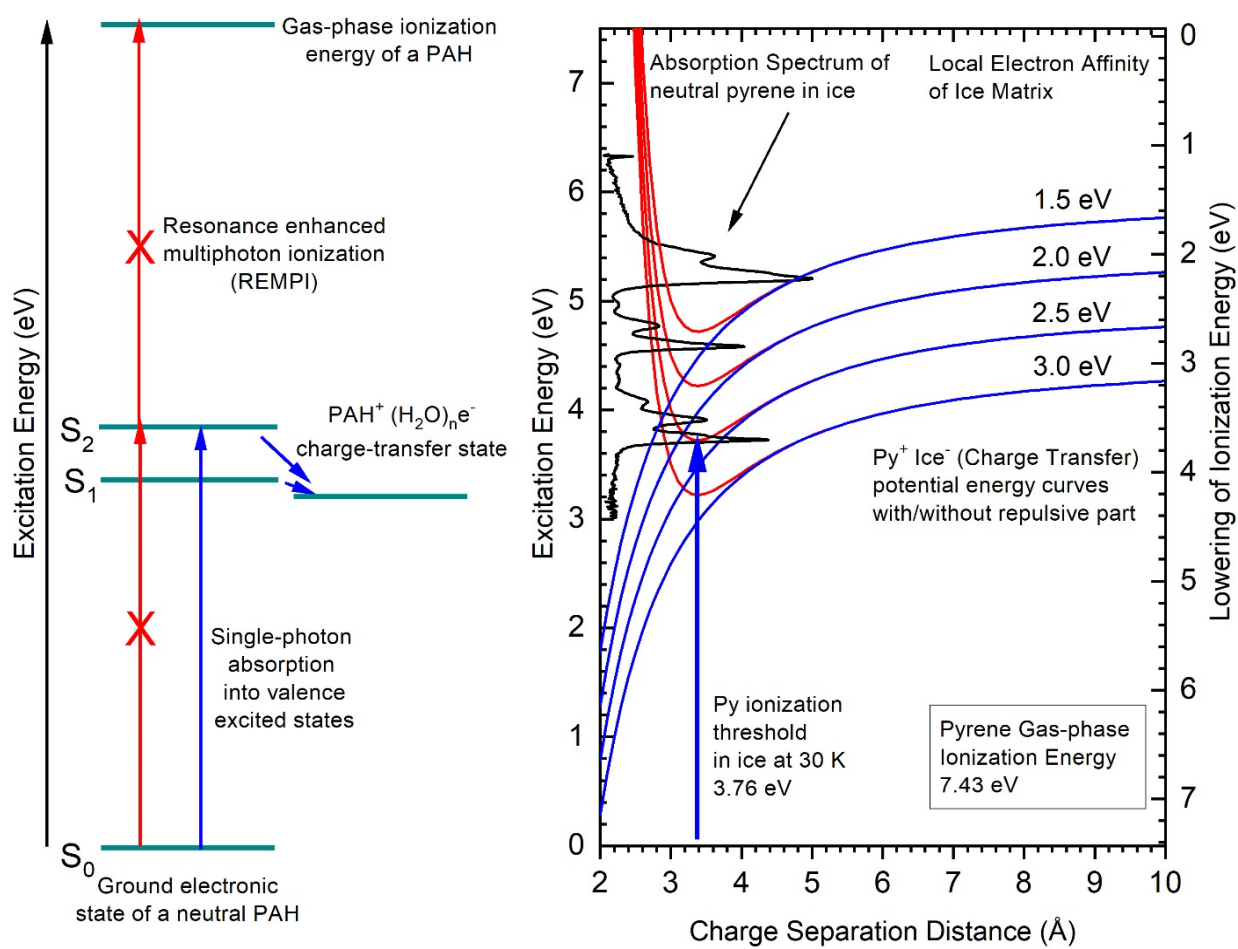

**Fig. 4. Electron-Ion Recombination in Ice:** Left Panel: Laser-assisted pumping between neutral and radical cation through ionization and ion-electron recombination of pyrene in amorphous ice at 5 K. Spectra from top to bottom: after deposition; photolysis at 355 nm increases radical cation absorption at 448 nm; photolysis at 448 nm depletes the radical cation, but increases the neutral absorption at 355 nm. Pumping between neutral and radical cation is continued by switching the laser wavelengths between 355 and 448 nm that results in hopping of an electron between the ice trap and pyrene molecule. Right Panel: Thermally induced ion-electron recombination of pyrene radical cation and an electron to produce neutral pyrene. Upper spectrum is the difference absorbance spectrum after ionization of neutral pyrene (depletion of neutral absorbance and increase of radical cation absorbance). During warm-up, pyrene radical cation is depleted and corresponding neutral pyrene is produced, demonstrating the availability of loosely bound electrons in the ice that recombine with pyrene radical cation. We note that amorphous ices are irreversibly transformed into crystalline ices at ~140 K and above.[21] The data at 60 K and 100 K is still from amorphous ice, indicating electron mobility even at these temperatures.

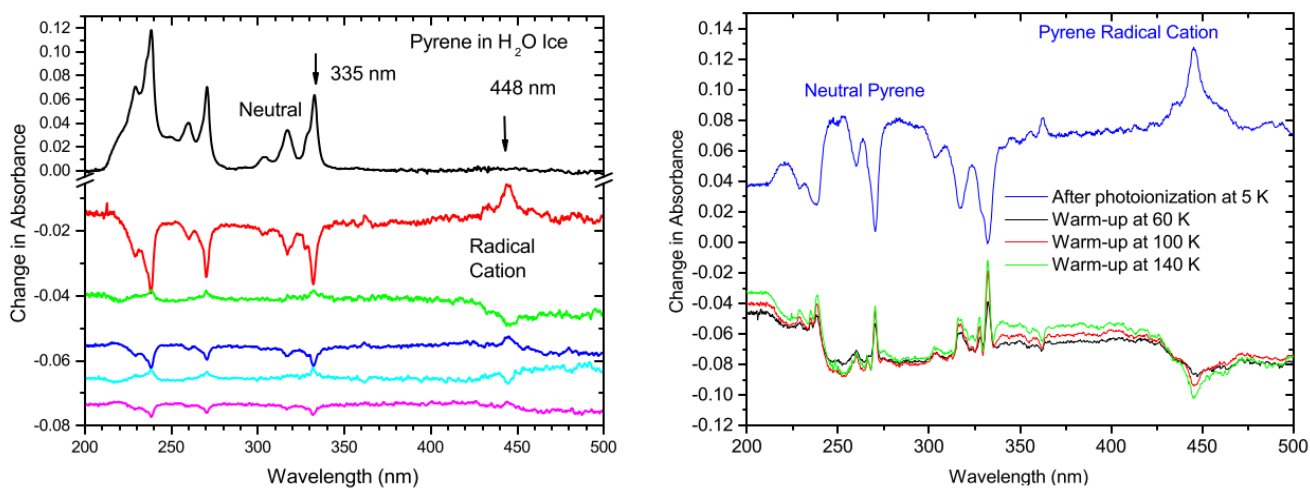

**Supplementary Materials**

**Supplementary Figure S5:** Wavelength Dependent Photoionization: Wavelength-dependent photoionization of perylene trapped in amorphous water ice at 5 K demonstrates involvement of initial valence excitation in the ionization process. Similar results were obtained with other PAHs used in this study. The y-axis presents the integrated absorbance of perylene cation band at that is appearing during the photoionization. Insert: Numbers with arrows show the sequence and wavelengths at which the ice was irradiated with a tunable laser.

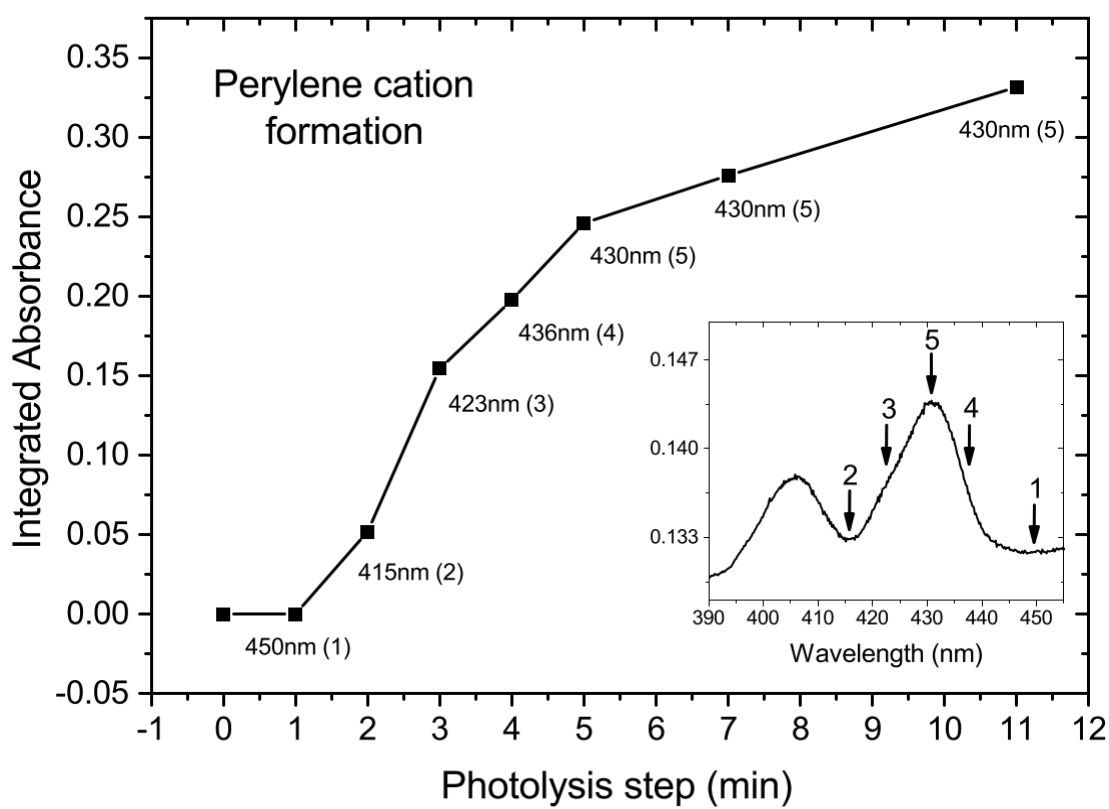

**Supplementary Figure S6:** Proof of Single-Photon Ionization: Photoionization of tetracene in water ice at 5 K at different cumulative laser doses beginning with 10 mW laser flux. Higher flux data was converted to a time-equivalent of the 10 mW dose (20 mW for 10 min = 10 mW for 20 min). Resulting data points fall on a single exponential curve (please see the red trace and fit) confirming that increasing laser energy increases photoionization yields (depletion of tetracene) proportionally, not quadratically - confirming a single-photon mediated photoionization process. The depletion of tetracene is obtained as degrease of an integrated absorption intensity over the strongest obtained absorption at ~280 nm.

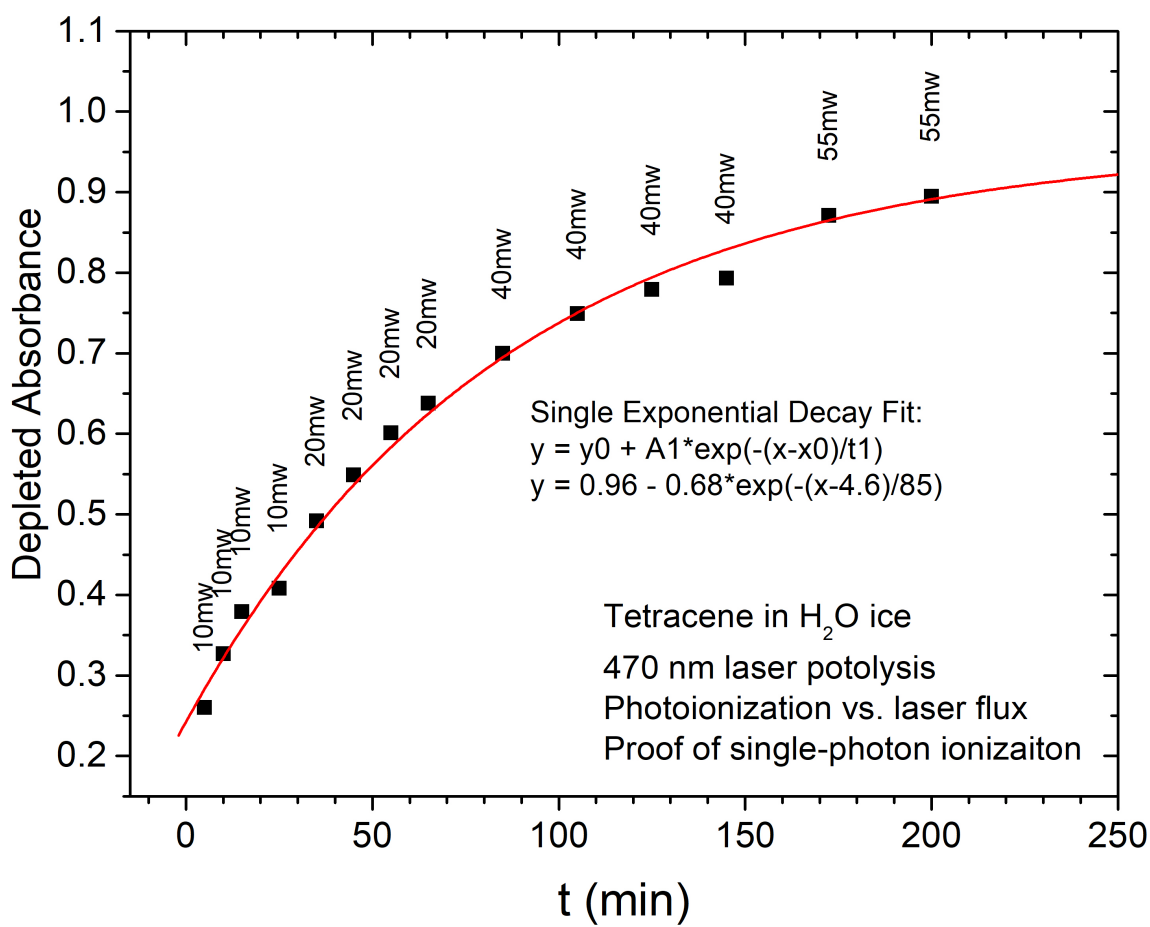

**Supplementary Figure S7:** Photoionization with UV-VIS Xe-Lamp. Bottom: UV absorption spectrum of neutral pyrene in water-ice (lower panel, black solid line), overlaid with the spectral intensity of Xe-arc lamp and a UV filter (arbitrary units). This combination ensures that pyrene molecules are only photo-excited into both the first ($S_1$, dark state) and second excited singlet state ($S_2$) that absorbs between 350 and 300 nm.[29, 49] It also ensures that pyrene radical cation that is generated during the photoionization is not further subjected to photodegradation (cutoff at >400 nm). Top: Difference absorption spectra during photolysis at different time intervals. These spectra clearly demonstrate that neutral pyrene in ice is depleted, with simultaneous appearance of pyrene radical cations.

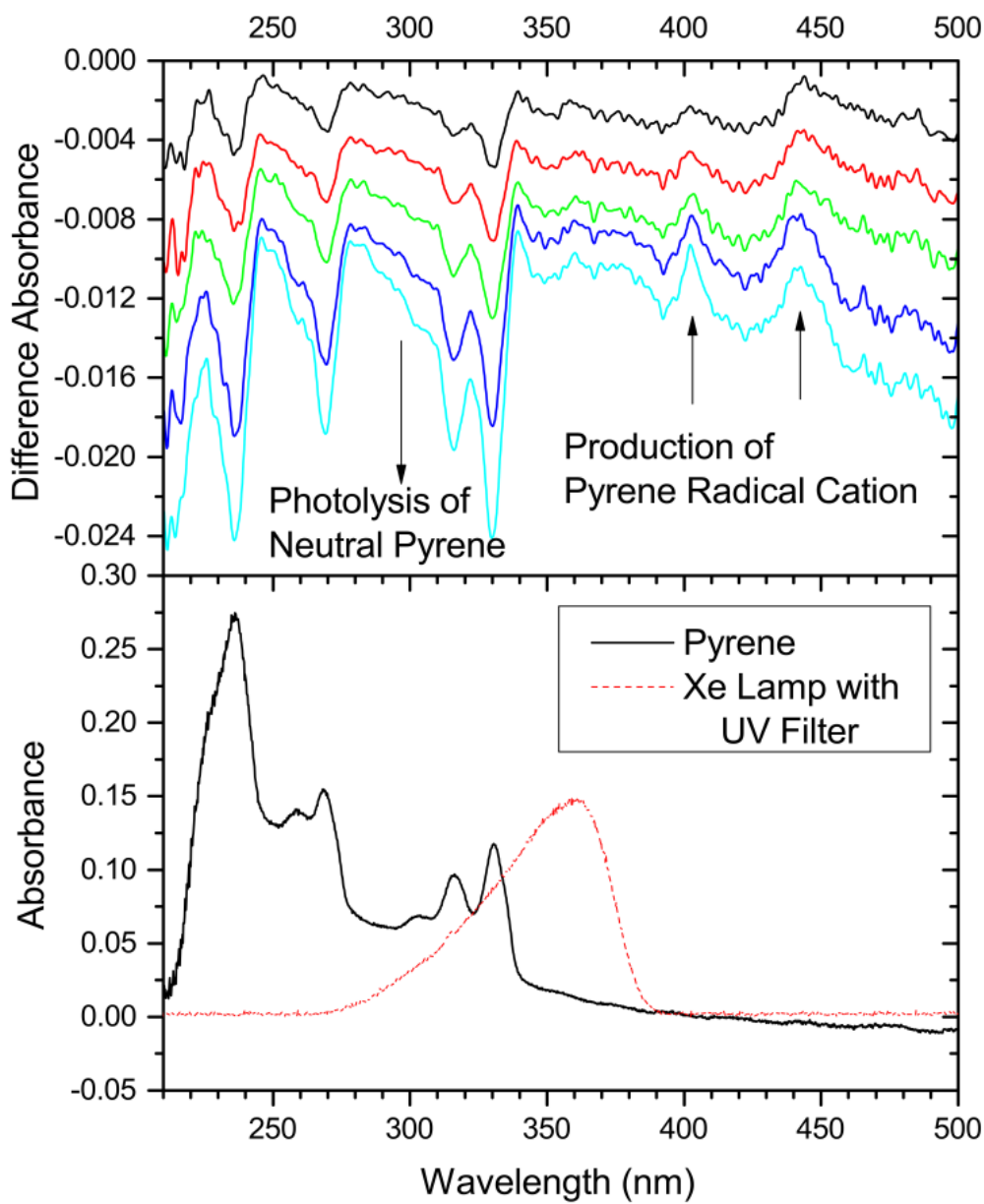

**Supplementary Figure S8:** Spectrometer lamp photoionization of tetracene in water-ice. Tetracene, the fourth member of the linear acene family after benzene, naphthalene, and anthracene, has characteristic strong UV absorption around 280 nm, and weak absorption starting at 480 nm, as seen in the top part of the figure. We have shown in Figure 1 that laser excitation into the longer wavelength absorption maximum around 475 nm results in photoionization of tetracene in water-ice. The data presented here demonstrates that PAH photoionization occurs even when exposed to an extremely weak light source such as a deuterium-halogen lamp that is used for measuring the absorption spectra. Further experiments that are not shown here revealed that photoionization under lamp conditions occurs more readily when the deuterium light was used (220-350 nm). We also obtained similar results when Xe-arc lamp was used with a monochromator that was tuned to the absorption maximum at 270 nm. All these studies demonstrate that photoionization of PAHs trapped in water-ice occurs with visible photons.

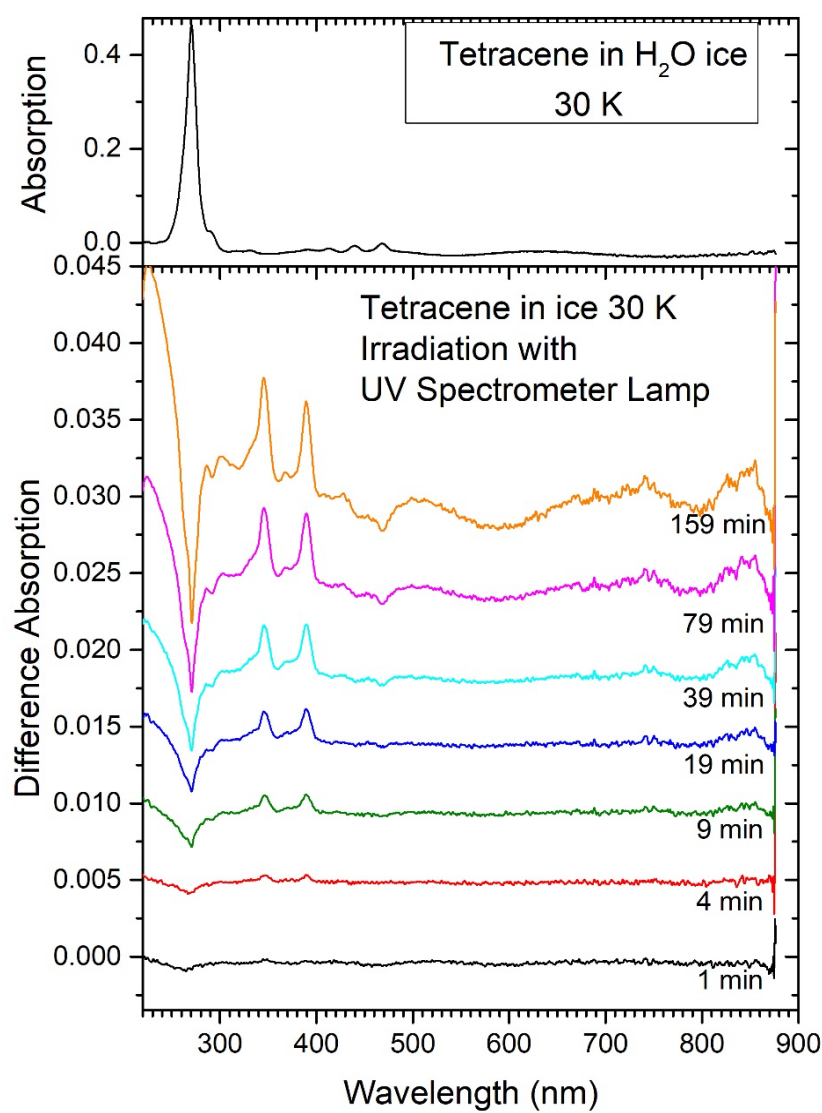

**Supplementary Figure S9:** Xe-arc-lamp photolysis of tetracene in Ar matrix. This figure shows that during photolysis of tetracene trapped in argon (Ar) matrix, which was prepared under similar conditions as the ice work shown in Figure S8 and no photoionization is detected. It is required to have an electron scavenger (acceptors) in the immediate vicinity in order to photoionize PAHs trapped in Ar matrices. In the case of water-ice, the water molecules act as the electron scavengers, increasing ionization efficiency of PAHs in ice to close to unity. Increase in UV absorption around 270 nm is due to the condensation of residual tetracene in the chamber (subsequent to deposition).

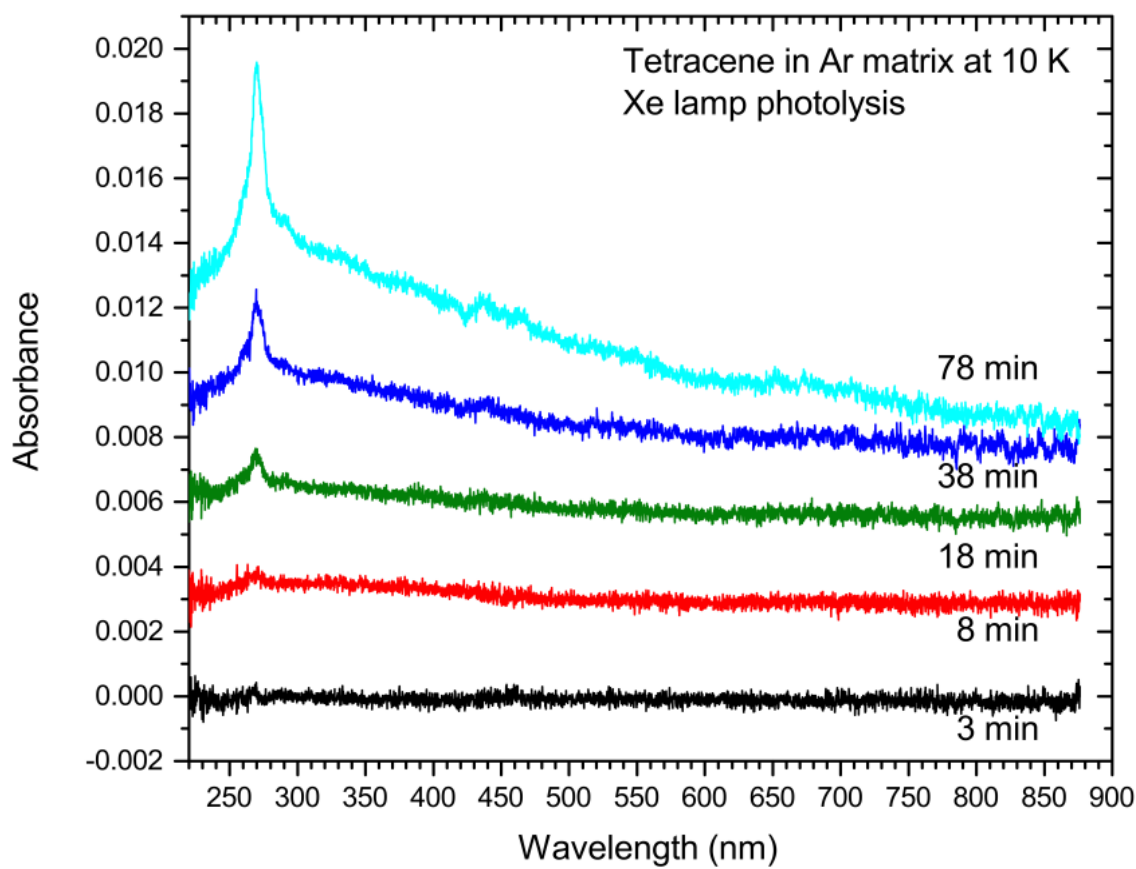

**Supplementary Figure S10:** Wavelength Dependence of Photoionization: Photoionization depletion of absorption from tetracene in water ice at 5 K at different wavelengths. Initial irradiation was conducted at 470 nm (the 0-0 vibronic $S_0$-$S_1$ transition)[50] as shown in the lower curve, reaching a plateau after about 125 min irradiation with a 55 mW laser. Tuning the laser wavelength to the higher vibronic transition (0-1 of the $S_0$-$S_1$ transition at 440 nm), photoionization continues further. However, production of the radical cation does not map on to the depletion of the neutral species due to further photochemistry of the radical cation. The absorption intensity is integrated over the strongest obtained tetracene absorption band at ~280 nm."

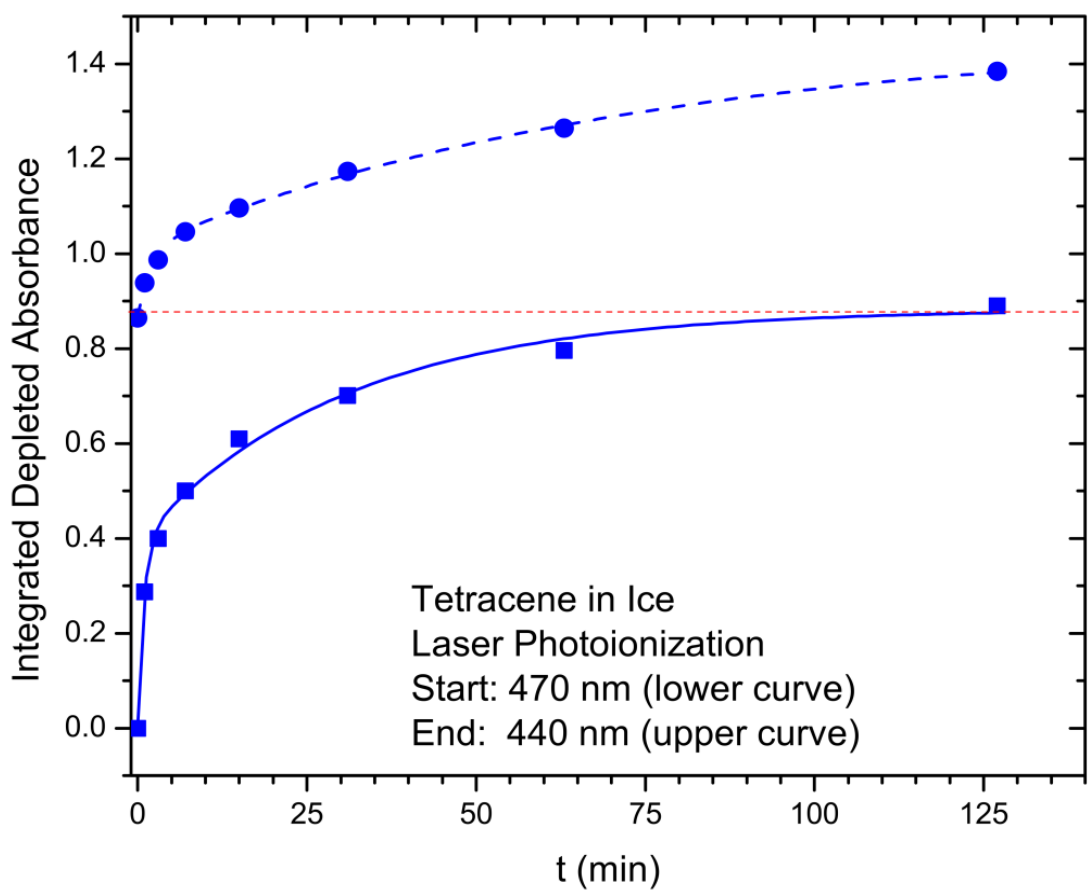

**Supplementary Figure S11:** Wavelength-Dependent Photoionization in Ice: Photoionization depletion of absorption from pyrene in water ice at 5 K at different wavelengths from two sets of experiments. Bottom curve (in blue) was taken when the ice was irradiated at one single wavelength, the 0-0 transition of the second excited singlet state ($S_2$) at 335 nm (see the absorption spectrum in the insert, $S_0$-$S_1$ absorption being dark).[29, 49] A monoexponential depletion and saturation at around 25% is observed. A slight decrease in the radical cation absorbance is noticed at prolonged photolysis times, indicating secondary reactions of the radical cation. In a second experiment under similar conditions, initially 335 nm photolysis was carried out for 1500 s. Subsequently, ice was photolyzed at various absorption maxima that covered vibronic bands of $S_2$ (320 nm and 306 nm) and higher excited states (273 nm and 241 nm).[29] A substantial depletion of neutral pyrene is observed each time as the wavelength was increased to probe absorption bands at higher energy. However, as noted in Figure S10 in the case of tetracene, the production of the radical cation does not map on to the depletion of the neutral species due to further photochemistry of radical cation.

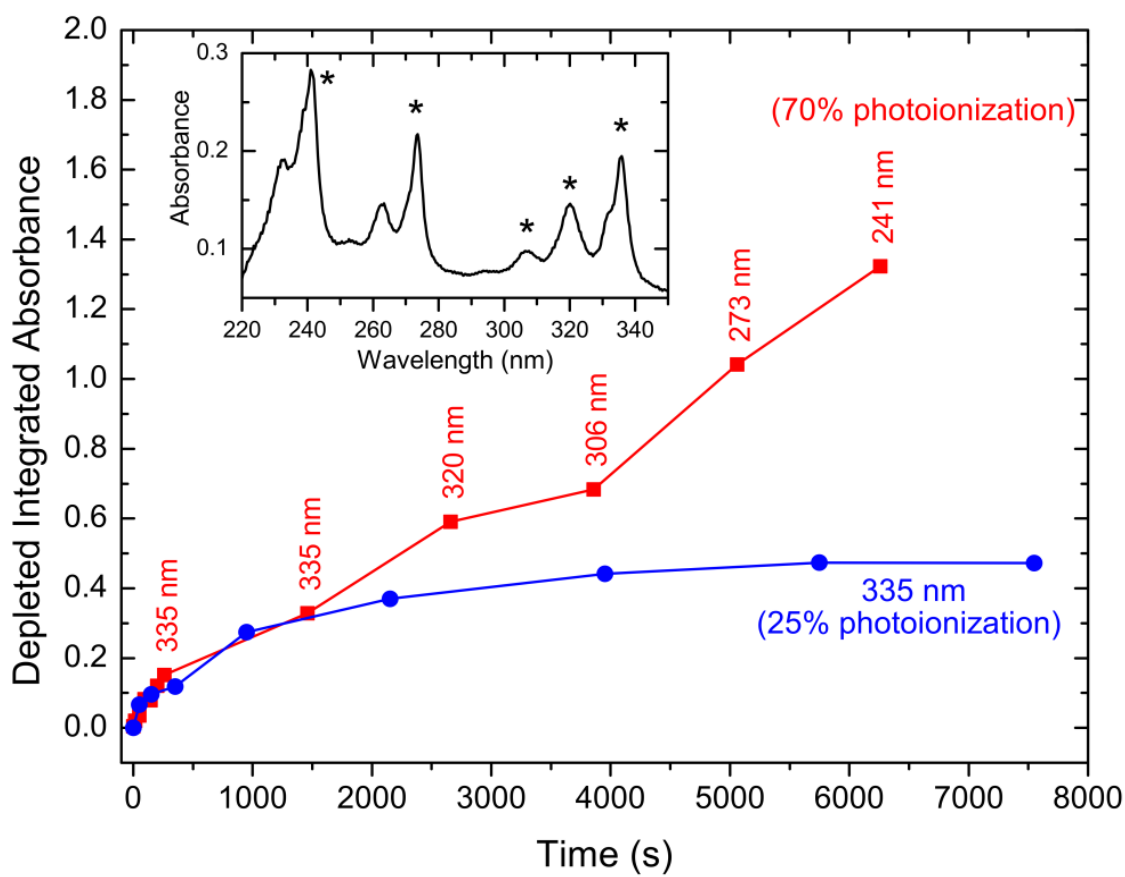

**Supplementary Figure S12**. FTIR spectrum of pyrene trapped in an amorphous water ice matrix. Black trace shows the overall absorption spectrum where the water ice absorptions are dominating. In the red trace, we have subtracted the water ice absorptions leaving only the pyrene bands visible (multiplied by the factor of five).

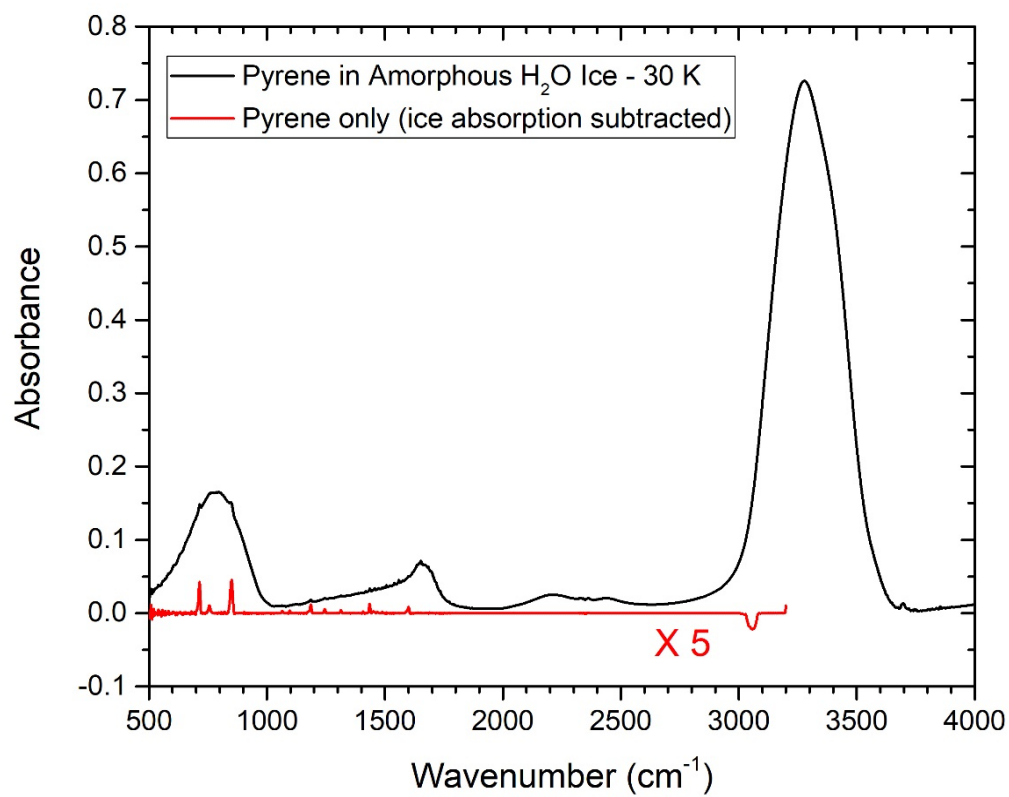